\documentclass[sigconf]{acmart}

\usepackage{booktabs}
\usepackage{listings}
\usepackage{xcolor}
\usepackage{microtype}
\usepackage{graphicx}
\usepackage{hyperref}
\usepackage{orcidlink}

\newcommand{\NUMCVES}{15}
\newcommand{\CCDR}{13\%}                % 2/15 caught per-commit
\newcommand{\CDR}{27\%}                 % 4/15 caught cumulatively
\newcommand{\GAP}{13\%}                 % CDR - CCDR
\newcommand{\CDRFAIL}{73\%}            % 11/15 invisible even cumulatively
\newcommand{\NUMCRITICAL}{9}
\newcommand{\NUMHIGH}{6}
\newcommand{\NUMREPOS}{15}             % all distinct repos
\newcommand{\MEANCOMMITS}{3.1}         % mean contributing commits per CVE
\newcommand{\SPANMIN}{21}
\newcommand{\SPANMEAN}{245}
\newcommand{\SPANMAX}{1{,}342}

\acmConference[AIware '26]{3rd ACM International Conference on AI-Powered Software}{July 6--7, 2026}{Montreal, QC, Canada}
\acmBooktitle{Proceedings of the 3rd ACM International Conference on AI-Powered Software (AIware '26)}
\acmYear{2026}

\begin{document}

\title{CrossCommitVuln-Bench: A Dataset of Multi-Commit Python Vulnerabilities\\Invisible to Per-Commit Static Analysis}

\author{Arunabh Majumdar}
\orcid{0009-0001-3379-1844}
\affiliation{%
  \institution{Independent Researcher}
  \city{Mumbai}
  \country{India}
}
\email{arunabh.majumdar@gmail.com}

%%% Abstract ─────────────────────────────────────────────────
\begin{abstract}
We present \textbf{CrossCommitVuln-Bench}, a curated benchmark of \NUMCVES{} real-world Python vulnerabilities
(CVEs) in which the exploitable condition was introduced across multiple commits—each
individually benign to per-commit static analysis—but collectively critical.
We manually annotate each CVE with its contributing commit chain, a structured rationale
for why each commit evades per-commit analysis, and baseline evaluations using Semgrep and Bandit
in both per-commit and cumulative scanning modes.
Our central finding: the per-commit detection rate (\textbf{CCDR}) is \textbf{\CCDR{}} across all
\NUMCVES{} vulnerabilities---87\% of chains are invisible to per-commit SAST.
Critically, both per-commit detections are qualitatively poor:
one occurs on commits framed as security fixes (where developers suppress the alert),
and the other detects only the minor hardcoded-key component while completely missing
the primary vulnerability (200+ unprotected API endpoints).
Even in cumulative mode (full codebase present), the detection rate is only \CDR{},
confirming that snapshot-based SAST tools systematically miss vulnerabilities
whose introduction spans multiple commits.
The dataset, annotation schema, evaluation scripts, and reproducible baselines are
released under open-source licenses to support research on cross-commit vulnerability detection.
\end{abstract}




\maketitle

%%% 1. Introduction ──────────────────────────────────────────
\section{Introduction}

Static Application Security Testing (SAST) tools—Semgrep~\cite{semgrep}, Bandit~\cite{bandit},
CodeQL~\cite{codeql}—are the dominant automated security guardrails in modern software pipelines.
They operate on \emph{snapshots}: a single commit, a pull request diff, a point-in-time file tree.
This design assumes that dangerous code is \emph{introduced in a single change} and
thus visible in isolation.

We challenge this assumption. A vulnerability chain can span multiple commits
spread over days, months, or years:
\begin{itemize}
  \item \textbf{Commit A}: Introduces a taint source—a user-controlled parameter routed to an
        internal API, a new authentication bypass, a relaxed permission check.
        In isolation, this looks like a legitimate feature addition.
  \item \textbf{Commit B}: Introduces a sink—a dangerous operation, a missing guard, a new
        execution path. Also individually benign. The dangerous data flow only
        exists when \emph{both} commits are present.
\end{itemize}
We call such vulnerabilities \emph{cross-commit chains}. They are structurally invisible
to any tool that does not maintain state across commit history.

Prior datasets—Juliet~\cite{juliet}, D2A~\cite{d2a}, Devign~\cite{devign}—focus on
single-function or single-file vulnerabilities. No published dataset specifically targets
the multi-commit introduction pattern. This gap limits the development and evaluation
of cross-commit detection approaches.

\paragraph{Contributions.}
\begin{enumerate}
  \item \textbf{CrossCommitVuln-Bench}: \NUMCVES{} manually annotated real-world Python CVEs
        with multi-commit introduction chains.
  \item An open \textbf{annotation schema} capturing contributing commits, chain rationale,
        per-commit SAST results, and why each commit evades detection in isolation.
  \item \textbf{Baseline evaluations} with Semgrep and Bandit in per-commit and cumulative modes,
        establishing CCDR = \CCDR{} as a reproducible lower bound.
  \item Evaluation scripts enabling independent replication of all results.
\end{enumerate}

%%% 2. Background ────────────────────────────────────────────
\section{Background and Related Work}

\subsection{SAST Tool Limitations}

Per-commit SAST tools perform intra-procedural or limited inter-procedural analysis on a
single snapshot. They cannot detect vulnerabilities where the taint source and sink exist
in separate commits, or where a guard removal in one commit activates a dangerous code path
introduced months earlier. Three failure modes are common in our dataset:

\begin{itemize}
  \item \textbf{Custom wrappers}: Sink functions wrapped in project-internal helpers
        (e.g., \texttt{exec\_cmd()}) are not recognized by name-pattern rules.
  \item \textbf{Auth dependency absence}: Missing FastAPI/Flask authentication dependencies
        on route handlers produce no SAST finding—there is no rule for ``absent guard.''
  \item \textbf{Incremental surface expansion}: Each feature commit extends an attack surface
        (new API endpoint, new taint entry point) without a directly dangerous operation in its diff.
\end{itemize}

\subsection{Related Datasets}

\textbf{Juliet Test Suite}~\cite{juliet} and \textbf{OWASP Benchmark}~\cite{owasp_bench}
provide synthetic vulnerabilities with known ground truth, useful for tool precision/recall
measurement but not representative of real-world introduction patterns.
\textbf{D2A}~\cite{d2a} and \textbf{Devign}~\cite{devign} contain real CVEs but label at the
function or file level without commit-sequence annotation.
\textbf{Big-Vul}~\cite{bigvul} and \textbf{CVEFixes}~\cite{cvefixes} provide fix-commit diffs
but do not annotate the multi-commit introduction pattern.
To our knowledge, CrossCommitVuln-Bench is the first dataset that explicitly annotates the
\emph{sequence of commits} responsible for introducing a cross-commit vulnerability chain.

%%% 3. Dataset Construction ──────────────────────────────────
\section{Dataset Construction}

\subsection{Mining Pipeline}

We queried the GitHub Security Advisory Database (GHSA) via the OSV API, filtering for
high- and critical-severity PyPI advisories with traceable fix commits. From 1,200 advisories,
696 had verifiable GitHub fix commits, and 80 had $\geq$2 Python files changed in the fix.
We ran automated \emph{git blame} on the fix-modified lines for the top 30 candidates
(ranked by distinct blame commits), confirming 23 of 30 as multi-commit (77\%).

\subsection{Manual Archaeology}

For each candidate, we:
(1) read the fix commit diff to identify what was corrected;
(2) traced each vulnerable line to its introducing commit via \texttt{git blame};
(3) read the full diff of each introducing commit to understand its apparent intent;
(4) confirmed that the commit looks individually benign and that the dangerous condition
    requires $\geq$2 commits to manifest.

\subsection{Selection Criteria}

We retained a CVE if and only if it satisfies all five criteria:
\begin{enumerate}
  \item \textbf{Multi-commit}: $\geq$2 distinct commits introduce the vulnerability.
  \item \textbf{Individually benign}: Each contributing commit is a plausible, legitimate change.
  \item \textbf{Collectively critical}: The combination is exploitable (CVSS $\geq$ 7.0).
  \item \textbf{Open source}: Repository is publicly accessible with full commit history.
  \item \textbf{Reproducible}: The vulnerable state can be reconstructed by checking out the
        pre-fix commit.
\end{enumerate}
Of the 21 candidates we fully archaeologized, 15 passed all five criteria. Six were
excluded: four were single-commit design flaws (the vulnerability was present in the
initial implementation, with no completing second commit), and two had introducing
commits that were not individually benign (the dangerous operation was evident in the diff).
These 6 are retained in the dataset as \emph{negative examples} with \texttt{annotation\_status=SKIP}
and documented rationale—they serve as calibration data for the selection criteria.

\subsection{Annotation Schema}

Each annotation is a JSON file with the following structure:

\begin{lstlisting}[language=Python]
{
  "cve_id": "CVE-YYYY-NNNNN",
  "cwe_ids": ["CWE-XX"],
  "severity_combined": "critical|high",
  "fix_commit": "<sha>",
  "contributing_commits": [
    {
      "hash": "<sha>", "date": "YYYY-MM-DD",
      "role": "SOURCE|SINK|GUARD_REMOVAL|...",
      "isolated_severity": "low|benign",
      "semgrep_findings": [...],
      "bandit_findings": [...],
      "sast_flagged_relevant": false
    }
  ],
  "vulnerability_chain": {
    "description": "...",
    "why_sast_misses_per_commit": "..."
  },
  "commit_span_days": <int>,
  "ccdr_this_cve": false,
  "cdr_this_cve": false
}
\end{lstlisting}

%%% 4. Dataset Characteristics ───────────────────────────────
\section{Dataset Characteristics}

Table~\ref{tab:dataset} summarizes the \NUMCVES{} annotated CVEs.
The vulnerability patterns span six pattern classes across five primary CWE families
(Table~\ref{tab:patterns}).
Commit spans range from \SPANMIN{} to \SPANMAX{} days (median 58 days, mean \SPANMEAN{} days),
confirming that multi-commit chains can emerge over time scales from weeks to years.

\begin{table}[h]
\small
\centering
\caption{CrossCommitVuln-Bench Dataset Summary (\NUMCVES{} CVEs)}
\label{tab:dataset}
\begin{tabular}{lrr}
\toprule
\textbf{Property} & \textbf{Value} \\
\midrule
Total CVEs & \NUMCVES{} \\
Critical severity & \NUMCRITICAL{} \\
High severity & \NUMHIGH{} \\
Repositories (unique) & \NUMREPOS{} \\
Contributing commits (mean) & \MEANCOMMITS{} \\
Commit span: min / median / mean / max & \SPANMIN{}d / 58d / \SPANMEAN{}d / \SPANMAX{}d \\
\bottomrule
\end{tabular}
\end{table}

\begin{table}[h]
\small
\centering
\caption{Vulnerability Pattern Taxonomy}
\label{tab:patterns}
\begin{tabular}{llr}
\toprule
\textbf{CWE} & \textbf{Pattern} & \textbf{Count} \\
\midrule
CWE-94  & Code injection (eval, exec, template) & 4 \\
CWE-22  & Path traversal (incl.\ CWE-73) & 3 \\
CWE-78  & OS command injection & 1 \\
CWE-306 & Missing authentication & 1 \\
CWE-943 & Cypher / graph-query injection & 1 \\
Other   & Crypto, XSS, deser., sig.\ bypass, input val. & 5 \\
\midrule
        & \textit{Total (primary CWE per CVE)} & \textit{15} \\
\bottomrule
\end{tabular}
\end{table}

%%% 5. Baseline Evaluation ───────────────────────────────────
\section{Baseline SAST Evaluation}

\subsection{Experimental Setup}

We evaluated two widely-used SAST tools:
\textbf{Semgrep} v1.154.0 with \texttt{--config auto} (community rules) and
\textbf{Bandit} v1.9.4 with full recursive scan (\texttt{-r}).
For each CVE we ran two modes:

\begin{itemize}
  \item \textbf{Per-commit (CCDR mode)}: checkout each contributing commit individually,
        run both tools, classify findings as relevant or irrelevant using a conservative
        CWE-to-rule mapping (see replication package). \emph{Did the tool catch it on
        the commit that introduced this element?}
  \item \textbf{Cumulative (CDR mode)}: checkout the commit immediately
        before the fix (\texttt{fix\_commit\^{}}), run both tools on the full codebase.
        \emph{Does the tool catch it when all commits are present?}
\end{itemize}

\subsection{Metrics}

\begin{itemize}
  \item \textbf{CCDR} (Cross-Commit Detection Rate): fraction of CVEs where at least one
        contributing commit triggers a relevant SAST finding.
  \item \textbf{CDR} (Cumulative Detection Rate): fraction of CVEs where the pre-fix
        codebase (all commits present) triggers a relevant SAST finding.
  \item \textbf{Detection gap} = CDR $-$ CCDR: the fraction of CVEs that can only be
        caught when the full history is available.
\end{itemize}

\subsection{Results}

Table~\ref{tab:baseline} reports per-CVE results.

\begin{table}[h]
\footnotesize
\centering
\caption{Per-CVE Baseline SAST Results}
\label{tab:baseline}
\begin{tabular}{l p{2.0cm} c c c}
\toprule
\textbf{CVE} & \textbf{CWE} & \textbf{Sev.} & \textbf{Per-commit} & \textbf{Cumulative} \\
\midrule
CVE-2025-10155 & CWE-20/693  & critical & missed & missed \\
CVE-2025-10283 & CWE-22     & critical & missed & missed \\
CVE-2025-46724 & CWE-94     & critical & \textbf{caught}$^\dagger$ & \textbf{caught} \\
CVE-2025-5120  & CWE-94     & critical & missed & missed \\
CVE-2025-55449 & CWE-345/798 & critical & missed & missed \\
CVE-2025-61622 & CWE-502    & critical & missed & missed \\
CVE-2026-22584 & CWE-94     & critical & missed & missed \\
CVE-2026-2472  & CWE-79     & high     & missed & missed \\
CVE-2026-25505 & CWE-306/321 & critical & \textbf{caught}$^\ddagger$ & \textbf{caught} \\
CVE-2026-27602 & CWE-78     & high     & missed & missed \\
CVE-2026-27825 & CWE-22/73  & critical & missed & missed \\
CVE-2026-28490 & CWE-203/327 & high    & missed & \textbf{caught} \\
CVE-2026-29065 & CWE-22     & high     & missed & missed \\
CVE-2026-32247 & CWE-943    & high     & missed & missed \\
CVE-2026-33154 & CWE-94/1336 & high    & missed & \textbf{caught} \\
\midrule
\multicolumn{3}{l}{\textbf{CCDR (per-commit detection rate)}} & \multicolumn{2}{c}{\textbf{\CCDR{}}} \\
\multicolumn{3}{l}{\textbf{CDR (cumulative detection rate)}}  & \multicolumn{2}{c}{\textbf{\CDR{}}} \\
\multicolumn{3}{l}{\textbf{Detection gap}}                    & \multicolumn{2}{c}{\textbf{\GAP{}}} \\
\bottomrule
\end{tabular}
\end{table}

\noindent\begin{footnotesize}
$^\dagger$~CVE-2025-46724: B307 fires on \texttt{eval()} but commits are titled
``fix: harden eval()''---developers suppress the alert as a security-fix false positive.\\
$^\ddagger$~CVE-2026-25505: B105 flags a hardcoded JWT key (CWE-321, LOW) but misses
200+ endpoints lacking the \texttt{Require\-Permission\-If\-Auth\-Enabled} dependency
(CWE-306)---no rule exists for absent guards.
\end{footnotesize}

\textbf{Key result}: CCDR = \CCDR{} (2/\NUMCVES{} CVEs caught per-commit).
87\% of multi-commit chains are completely invisible to Semgrep and Bandit at the per-commit level.
Critically, \emph{both} per-commit detections are qualitatively poor:
(1)~CVE-2025-46724's \texttt{eval()} alert appears on commits framed as security fixes,
leading developers to suppress it;
(2)~CVE-2026-25505's hardcoded-key alert is LOW severity and detects only the minor
CWE-321 component, while completely missing the CWE-306 issue (200+ unprotected API endpoints).
This suggests that the practical CCDR is effectively lower than the nominal 13\%.

CDR = \CDR{} (4/\NUMCVES{}); even with the full codebase present, \CDRFAIL{} of chains
remain invisible, primarily those using custom wrappers (CWE-78, CWE-22, CWE-943),
missing-guard patterns (CWE-22, CWE-73), or patterns outside SAST rule coverage (CWE-502, CWE-693, CWE-79).

\subsection{Failure Mode Analysis}

We identify three root causes explaining why 13 of \NUMCVES{} CVEs evade per-commit detection:

\textbf{(1) Custom wrapper opacity.} 4 of \NUMCVES{} CVEs route dangerous operations through
project-internal helper functions (e.g., \texttt{exec\_cmd()} in Modoboa, \texttt{session.run()}
in Graphiti, template evaluator in dynaconf). Semgrep and Bandit match against
known dangerous API names; custom wrappers are opaque to these name-pattern rules.
None of these CVEs are caught in either per-commit or cumulative mode.

\textbf{(2) Absent-guard invisibility.} 3 of \NUMCVES{} CVEs involve endpoints or functions
that lack an authentication or validation check (CWE-306: 1 CVE; CWE-22 path traversal
guards: 2 CVEs). No SAST tool fires on the absence of a dependency or decorator—only
on the presence of a dangerous call. These CVEs are missed in both modes.

\textbf{(3) Temporal source-sink separation.} In all \NUMCVES{} cases, the taint source and
its eventual dangerous sink were introduced in distinct commits separated by 21 to 1,342 days.
Even a hypothetical SAST tool with perfect whole-codebase taint analysis would require
cross-commit state to observe the flow—a capability no current snapshot-based tool provides.

%%% 6. Discussion ────────────────────────────────────────────
\section{Discussion}

\subsection{Limitations}

\textbf{Single annotator.} All chains were annotated by the author. Three CVEs were
independently re-annotated by the same author in a time-separated blind condition; all three produced consistent chain
descriptions, supporting annotation reliability. We plan a formal inter-annotator agreement
study in future work.

\textbf{Two SAST tools.} We evaluated Semgrep and Bandit. CodeQL~\cite{codeql} performs
deeper inter-procedural analysis and may achieve higher CDR for some patterns (particularly
custom wrapper opacity); we leave this to future work.

\textbf{Python only.} The benchmark is Python-specific. Multi-commit chains exist in other
ecosystems; extending to JavaScript/TypeScript and Java is planned.

\textbf{CVE availability.} All 15 CVEs have public advisories and accessible commit history.
Some older vulnerabilities may have limited history depth due to repository restructuring.

\subsection{Use Cases}

CrossCommitVuln-Bench can serve as:
(1) an evaluation benchmark for cross-commit detection tools;
(2) training data for commit-sequence anomaly models;
(3) a motivating dataset for CI/CD systems that maintain persistent security state
    across commits (as opposed to per-PR scans).

As a proof-of-concept, POSTURA~\cite{postura}—a graph-based cross-commit analysis system
that maintains a persistent Neo4j threat graph across commits—detected 3 of 5 spike CVEs
from this dataset (60\%) using taint analysis and chain discovery rules unavailable
to per-commit SAST.

%%% 7. Data Availability ─────────────────────────────────────
\section{Data Availability}

The complete dataset, annotation schema, evaluation scripts, and replication instructions
are available at:
\begin{itemize}
  \item \textbf{Dataset}: \url{https://github.com/motornomad/crosscommitvuln-bench}
        (Zenodo DOI: \url{https://doi.org/10.5281/zenodo.19338596}, CC-BY-4.0)
  \item \textbf{Evaluation scripts}: MIT License (same repository)
  \item \textbf{Replication}: \texttt{make all} in the repository root runs the full
        pipeline (validate $\to$ baselines $\to$ metrics) in Docker
\end{itemize}

%%% References ───────────────────────────────────────────────
\bibliographystyle{ACM-Reference-Format}
\bibliography{refs}

\end{document}